\documentclass[sigconf,natbib=false,nonacm]{acmart}

\AtBeginDocument{%
  }

\RequirePackage[
  datamodel=acmdatamodel,
  style=acmnumeric,
  ]{biblatex}
\usepackage{amsmath}

\usepackage{booktabs}
\usepackage{multirow}

\usepackage{algorithm}
\usepackage{algorithmic}

\usepackage{enumitem}

\begin{document}

\title{OxygenREC-v2: Internalizing Discrimination into Generative Recommendation}

\author{Guo Tang$^*$, Hanye Wu$^*$, Changjiang Han$^*$, Qingyang Li, Ming Zhang$\dagger$, Xiangyu Qian, Yanchen Qiao, Huanjie Wang, Zhi Ma, Zhen Li, Yaqiang Zang$\dagger$, Pinghua Gong$\dagger$}
\affiliation{%
  \institution{JD.COM, Beijing, China}
  \country{\{tangguo.26, wuhanye.1, hanchangjiang.rand, liqingyang.alex, zhangming229, qianxiangyu3, qiaoyanchen1, wanghuanjie.cielo, mazhi6, lizhen444, zangyaqiang.1, gongpinghua1\}@jd.com}
}

\renewcommand{\shortauthors}{Tang et al.}

\begin{abstract}
Generative recommendation unifies retrieval and ranking within a single model by autoregressively decoding semantic identifier (SID) sequences.
Yet reliably incorporating behavior signals from clicks, cart additions, and orders remains challenging.
Existing approaches either jointly optimize generative and discriminative objectives, requiring delicate trade-offs, or use a separate ranker as a post-hoc reinforcement-learning reward, risking out-of-distribution scoring and reward misalignment.
We propose \textbf{OxygenREC-v2}, a generative recommender that \textbf{I}nternalizes \textbf{D}iscrimination into \textbf{G}enerative \textbf{R}ecommendation (\textbf{IDGR}).
Rather than adding a separate discriminative objective, OxygenREC-v2 uses logged behavior to condition generation and supervise training.
During pre-training, a behavior instruction conditions generation on the target behavior.
During post-training, future interaction behaviors are exploited as privileged knowledge in our entropy-aware trajectory optimization self-distillation framework, enabling reward-model-free policy optimization. Throughout both training stages, OxygenREC-v2 maintains a single unified backbone. 
We implement OxygenREC-v2 as a 3B-parameter, 1B-activated MoE and deploy it on JD.com's large-scale e-commerce platform.
Across multiple online A/B tests, OxygenREC-v2 improves user click-through conversion rate (UCTCVR) by 1.6--4.4\% and GMV by 2.8--6.8\% over OxygenREC-v1.
\end{abstract}



\maketitle

\let\thefootnote\relax\footnotetext{$^*$These authors contributed equally to this work.}
\let\thefootnote\relax\footnotetext{$^{\dagger}$Corresponding authors.}

\section{Introduction}
\label{sec:intro}
\begin{figure*}[t]
    \centering
    \includegraphics[width=\textwidth]{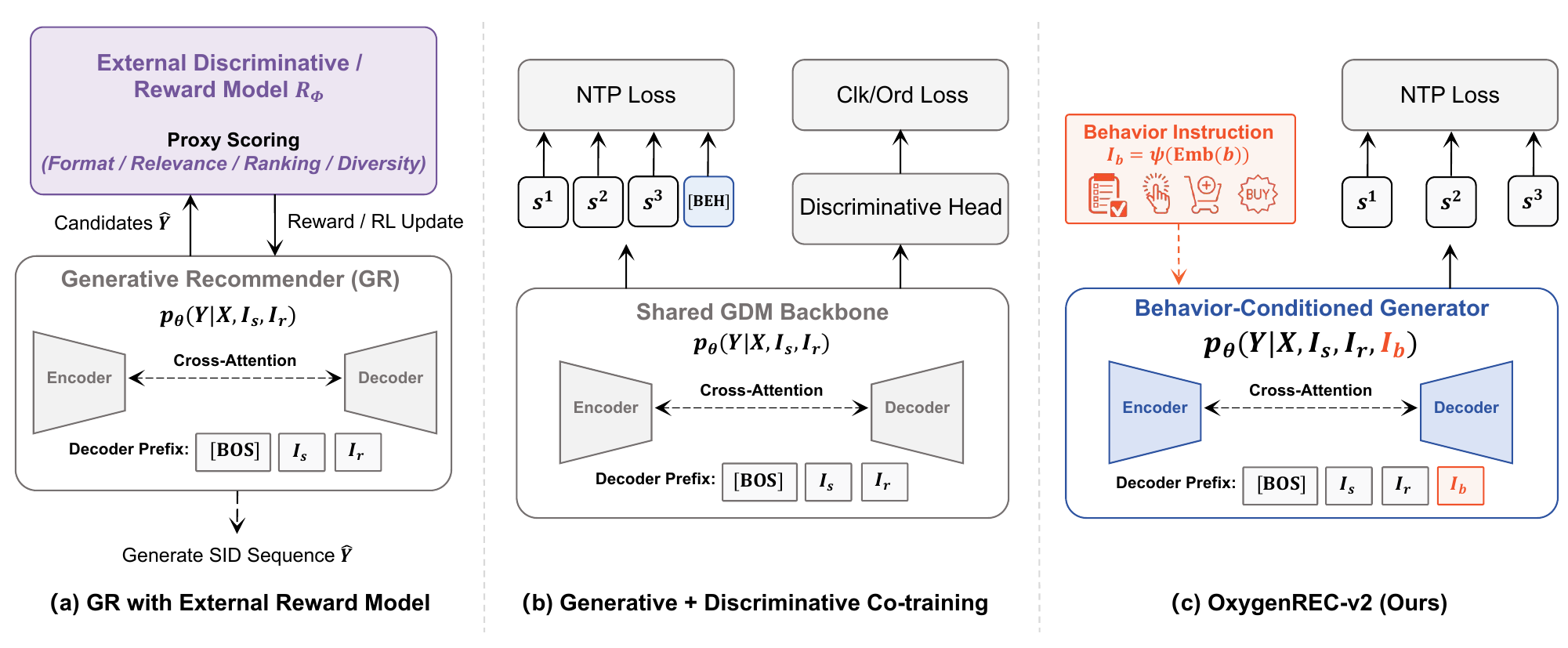}
    \caption{Three paradigms for coupling the behavior signal with a generative
recommender. \textbf{(a) External reward model:} a behavior-agnostic generator
is scored after generation by a separately trained reward model. \textbf{(b) Generative--discriminative
co-training:} a shared generative--discriminative model (GDM) backbone carries a
discriminative head that predicts behavior at the output end, but the
candidate distribution stays behavior-agnostic. \textbf{(c) OxygenREC-v2
(ours):} the behavior instruction $I_b$ enters the decoder prefix and conditions
generation from the first step, internalizing the behavior signal into the
generator.}
    \Description{A three-panel comparison of generative recommendation designs:
    an external reward model that scores generated candidates, a
    generative-discriminative co-training design with an output-side
    discriminative head, and our behavior-conditioned generator that places the
    behavior instruction in the decoder prefix.}
\label{fig:overview}
\end{figure*}

Generative Recommendation (GR) replaces the cascaded retrieval-and-ranking
pipeline of industrial recommenders with a single generative model that decodes
item identifiers directly from a user's context~\cite{tiger,deng2025onerec}.
Unifying the two stages into one decoding process removes their objective
mismatch, and recent GR systems match or surpass strong pipelined recommenders
at industrial scale~\cite{deng2025onerec,zhai2024hstu}. Yet these systems use the
discriminative feedback from clicks, cart-adds, and orders only indirectly. Most
inject it after generation, through a separately trained ranking model that
serves as a proxy reward for reinforcement learning~\cite{schulman2017ppo,ouyang2022instructgpt};
our industrial baseline, for instance, fuses format, relative-order, ranking, and
diversity rewards from such a model~\cite{oxygenrec} (Figure~\ref{fig:overview} (a)).

Two problems recur in this design. First, an offline ranking model scores
out-of-distribution outputs unreliably, so the policy can raise its proxy score
without improving recommendation quality, a form of reward
hacking~\cite{gao2023scaling,skalse2022defining}. Second, the competing reward
terms pull in different directions, and balancing their weights becomes its own
tuning problem. Both problems share one cause: during pre-training the generator
never sees the behavior it should promote. A click target and an order target
carry the same instruction, so the model returns the same candidate distribution
for both, and the behavior signal can enter only afterward. This raises a
question:

\begin{center}
\textit{\textbf{How can the discriminative behavior signal be
internalized directly into the generative model's training objective,
rather than delegated to a separate external reward model?}}
\end{center}

Placing behavior at the output end~\cite{ma2018esmm,tang2020ple} does not answer
it. A GR model can append a behavior token after the item identifier or predict
behavior from the final decoder state, letting generation and discrimination
share one backbone (Figure~\ref{fig:overview} (b)). Both still \emph{predict}
behavior after generation instead of using it to \emph{condition} generation, so
the candidate distribution is identical across target behaviors. Reranking
reorders the decoded items but cannot change which items are generated.

We instead condition generation on behavior directly. The behavior hierarchy
already in session logs, which orders exposure, click, add-to-cart, and order by
intent strength, is a supervision signal that runs through both training stages
without any extra model: a behavior label that \emph{conditions} generation in
pre-training is naturally preserved as the supervision signal throughout post-training (Figure~\ref{fig:overview} (c)). Making the signal internal rather
than external removes the source of the proxy-reward problem.


While behavior-aware pre-training internalizes user behavior into generation, enabling behavior-guided policy optimization during post-training without relying on external reward models remains challenging. Verifiable rewards provide an optimization objective without relying on external reward models, but only provide sparse supervision for behavior-aware policy optimization. Recent On-Policy Distillation (OPD)~\cite{opd} can provide token-level teacher-student distillation signals for behavior-aware policy update, but maintaining an external rapid iterative teacher is costly in recommendation systems~\cite{rec-distill}. A natural idea is to utilize user's future interactive behaviors as the privileged information for On-Policy Self-Distillation (OPSD)~\cite{opsd}. However, directly applying privileged self-distillation inevitably introduces privilege-dependent bias.

We propose \textbf{OxygenREC-v2}, a generative recommender that
\emph{\textbf{I}nternalizes \textbf{D}iscrimination into \textbf{G}enerative \textbf{R}ecommendation} (\textbf{IDGR}):
it folds the behavior signal into the generator's objective at both training
stages. In pre-training, a
\emph{behavior-aware instruction} $I_b$ encodes each target's ground-truth
behavior label and enters the decoder prefix, conditioning generation from the
first decoding step so the candidate distribution shifts with the target
behavior. 

In post-training, we use a unified \emph{\textbf{E}ntropy-\textbf{A}ware \textbf{T}rajectory \textbf{O}ptimization \textbf{S}elf-\textbf{D}istillation framework} (\textbf{EA-TOSD}). On the basis of the behavior-conditioned generator, we further utilize future user interaction behavior as teacher privileged knowledge for preference alignment, while employing an entropy-aware strategy to prevent privileged distillation bias.

On JD.com's industrial e-commerce platform, OxygenREC-v2 improves HR@1 from 4.62\%
to 5.64\% and HR@512 from 43.24\% to 44.14\% over the pre-trained baseline, and
leads a proxy-reward baseline on six of seven retrieval and ranking metrics, with
the largest gains on order, the sparsest and highest-value behavior. Deployed as
a 3B-parameter, 1B-activated (3B-A1B) MoE, it raises online User Click-Through
Conversion Rate (UCTCVR) by 1.6--4.4\% and GMV by 2.8--6.8\% across multiple scenarios.

\noindent Our contributions are summarized as follows:
\begin{itemize}[itemsep=2pt,topsep=2pt,parsep=0pt,leftmargin=*]
  \item \textbf{Behavior-conditioned pre-training.} The behavior-aware
  instruction $I_b$ conditions every decoding step on the target behavior, and
  behavior-aware training aligns the supervision by expanding, filtering, and
  value-weighting logged targets.

  \item \textbf{Entropy-aware self-distillation post-training framework.} We propose a unified reward-model-free post-training framework that directly internalizes user behavioral signals into policy optimization by jointly leveraging verifiable trajectory optimization, a teacher informed by the user's future behavior, and entropy-aware privileged distillation.
   
  \item \textbf{Industrial validation.} On JD.com's
    industrial e-commerce platform,
  OxygenREC-v2 improves offline retrieval and ranking compared to the
  baseline method and raises online UCTCVR and GMV in bucketed A/B tests.
\end{itemize}
\section{Related Work}
\label{sec:related}

\paragraph{Behavior signals in generative recommendation.}
Generative retrieval autoregressively decodes item identifiers from interaction histories~\cite{tiger,deng2025onerec}. Classical industrial recommenders model heterogeneous actions such as clicks and conversions with multi-task discriminators~\cite{ma2018esmm,tang2020ple}. Recent generative systems largely retain this separation: OneRec uses a learned reward model~\cite{zhou2025onerectechnicalreport}, OneMall an online ranker as reward~\cite{zhang2026onemall}, and GenRec a dense preference model with a relevance gate~\cite{zou2026genrec}. Such proxies densify feedback but invite reward overoptimization under policy shift~\cite{gao2023scaling,skalse2022defining}. A complementary line internalizes behavior: MBGen jointly generates behavior and item tokens~\cite{liu2024mbgen}, while PinRec conditions generation on desired engagement outcomes~\cite{botta2026pinrec}. These studies establish behavior-conditioned generation; we study its reliable post-training without an external behavior scorer.

\paragraph{Verifiable optimization and privileged on-policy distillation.}
Recommender policies have long learned from logged feedback, including production-scale off-policy reinforce~\cite{chen2019topk}. RL with verifiable rewards instead anchors updates to ground truth checks, avoiding a learned judge's proxy error~\cite{tulu3,deepseekr1}. OneRec-V2 likewise optimizes real user feedback, reducing reward-hacking risk but retaining sparse positives and difficult multi-objective credit assignment~\cite{zhou2025onerecv2technicalreport}. Distillation supplies denser supervision. Beyond online behavior classifiers, P5 encodes tasks and preferences in prompts~\cite{geng2022p5}, PFD transfers post-event privileged features~\cite{xu2020pfd}, and AFE distills future interactions and item taxonomy~\cite{xie2022afe}. For Large Language Models (LLMs), OPD supervises student rollouts with dense token distributions~\cite{gu2024minillm,agarwal2024gkd,opd}; OPSD instead uses the same backbone but with privileged inputs~\cite{opsd}. Yet privileged teachers can reinforce unsupported reasoning, motivating purification and entropy-based selection~\cite{shen2026purified,eopd}. We adapt this line to generative recommendation by combining verifiable trajectory optimization with entropy-aware privileged self-distillation.

\section{Preliminaries}
\label{sec:preliminaries}

\paragraph{Generative Recommendation Setup.}
\label{sec:gr-setup}

We adopt an encoder--decoder generative recommender~\cite{tiger,deng2025onerec}.
The encoder maps a user's interaction context $X$ (behavioral sequences,
profile features, and contextual signals) into a sequence of hidden states.
The decoder then produces a recommendation autoregressively as a string of
semantic identifier (SID) tokens. Each item is tokenized into an ordered triple
$(s^1, s^2, s^3)$ drawn from three hierarchical codebooks via residual
quantization~\cite{lee2022rqvae}, so decoding one item takes three steps. We
write a decoded recommendation as a flattened SID-token sequence
$Y=(y_1,\dots,y_L)$, where each $y_t$ is a single SID token and $L$ is the number
of decoding steps; a single item corresponds to $L=3$.
Following the instruction-following design of the industrial system we build
on~\cite{oxygenrec}, the decoder is conditioned on a composite instruction
prompt $P = (I_s, I_r)$, where $I_s$ is the \emph{scenario instruction}
(encoding scene context and an optional trigger item) and $I_r$ is the
\emph{contextual reasoning instruction}, a dense embedding synthesized by a
nearline large language model~\cite{ouyang2022instructgpt} from user intent
signals. The pre-training objective maximizes the likelihood of the target SID
sequence $Y$ given the user context and instructions:
\begin{equation}
  p_\theta(Y \mid X, I_s, I_r).
  \label{eq:base}
\end{equation}
This setup defines the surface on which our framework operates: the
instruction prompt is the model's only input-side handle for shaping
generation, and the SID sequence is its only output.

\paragraph{The External Discriminator Pattern.}
\label{sec:external-discriminator}

The pre-training objective in Eq.~\eqref{eq:base} reflects what items a user
engaged with, but not the \emph{type} of engagement. Discriminative
information about behavior is conventionally injected after pre-training
through an external module. A separately trained ranking model
$R_\phi(\hat{Y}, X)$ scores the generated candidates $\hat{Y}$, and the policy
is fine-tuned by reinforcement learning against a fusion of proxy reward terms,
\begin{equation}
  \mathcal{R}(\hat{Y}, X) = \sum_{j} \alpha_j \, r_j(\hat{Y}, X),
  \label{eq:proxyreward}
\end{equation}
where each $r_j$ targets a different objective (format, ranking, diversity,
relevance) and the weights $\alpha_j$ are set by hand. We refer to this
two-stage arrangement, in which a pre-trained generator is paired with an
externally trained discriminator that supplies a proxy reward, as the
\emph{external discriminator pattern}. It is the dominant recipe in
industrial GR systems~\cite{oxygenrec} and the baseline against which our
framework is positioned. 

\section{Methodology}
\label{sec:method}
 \begin{figure*}[t]
      \centering
      \includegraphics[width=0.8\textwidth]
      {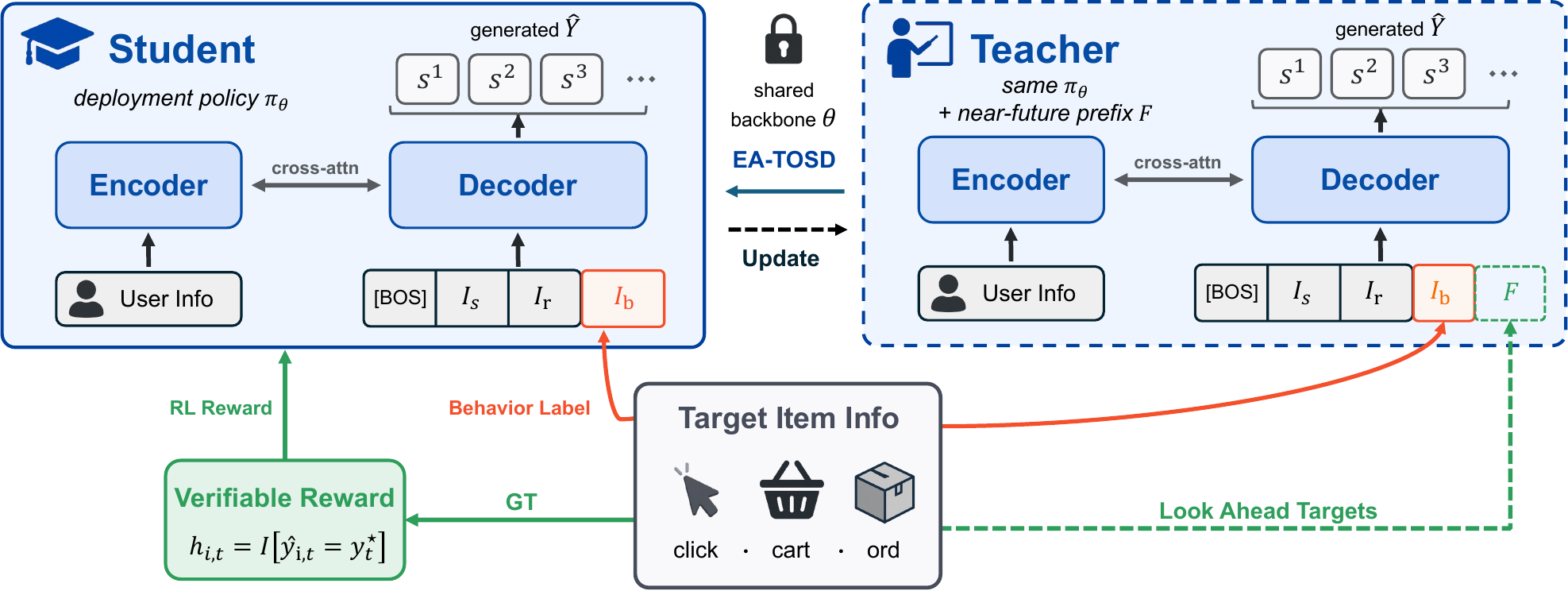}
      \caption{Overview of OxygenREC-v2. The behavior label of each logged target is embedded as the instruction $I_b$ and appended to the decoder prefix, so the deployed policy conditions SID generation on the target behavior. In post-training, the same target supplies behavior-filtered ground truth for the verifiable reward, while a training-only privileged teacher reuses the backbone with a near-future prefix $F$ to transfer future-informed preferences to the student.}
      \label{fig:frame-overview}
  \end{figure*}

Building on the generative recommender of Section~\ref{sec:preliminaries}, our
method adds the behavior signal to the training objective through four components.
The behavior-aware instruction conditions generation on the target behavior,
and behavior-aware training reweights the supervision toward informative
behaviors. Post-training then replaces the proxy ranking reward with a
verifiable behavior match and distills a privileged teacher that looks ahead to
the user's future behavior. A single behavior hierarchy, ordered by intent
strength from exposure through click and add-to-cart to order, runs through all
four: it sets the instruction priority, the supervised loss weights, and the
teacher's behavior threshold. Figure~\ref{fig:frame-overview} shows how the same
signal is reused: it conditions the deployed student through $I_b$, grounds the
verifiable reward against observed targets, and filters the near-future context
available only to the privileged teacher.

\subsection{Behavior-Aware Instruction}
\label{sec:behavior-instruct}

The scene and reasoning instructions specify where recommendations are made and to whom,
but not the behavior the recommendation should drive. The same user on a home feed,
where the goal is a click, and on the cart page, where the goal is an order, should
receive different candidates, yet without a behavior signal the model reads
identical instructions and returns the same distribution. Behavior is the axis
along which recommendation objectives differ: a click target favors exploration
and diversity; an order target favors high-conversion items. Expressing that
difference only through an external model that scores candidates afterward is
indirect and leaves the generator itself behavior-agnostic.

\paragraph{Behavior instruction $I_b$.}
We encode the target behavior $b$, one of click, cart, or order,
as an instruction $I_b$ and prepend it to the decoder. The behavior maps to a
reserved token whose embedding is read from the decoder's shared embedding table
and passed through a two-layer feed-forward projection $\psi$, giving
$I_b = \psi(\mathrm{Emb}(b))$. We concatenate $I_b$ after the scene and reasoning instructions, so
the decoder prefix becomes $[\textsc{bos}, I_s, I_r, I_b]$. At training time $b$ is the
behavior of the logged target, selected according to the priority order
$\text{order} \succ \text{cart} \succ \text{click} \succ \text{exposure}$; at
inference time $b$ is set by the business goal of the scene, with no auxiliary
model involved. The decoder attends to $I_b$ at every step, so the target behavior
shapes the whole autoregressive generation, most strongly at the first codebook
level $s^1$ that fixes an item's broad category. This yields the behavior-conditioned
objective
\begin{equation}
  p_\theta(Y \mid X, I_s, I_r, I_b), \qquad I_b = \psi(\mathrm{Emb}(b)).
\end{equation}

\subsection{Behavior-Aware Training}
\label{sec:behavior-training}

A user session logs many behavior events, but training on only a few targets per
sample discards much of the available supervision, and frequent exposures dominate
the rare cart and order events that carry the most value. This works against the
behavior instruction: $I_b$ names a target behavior, while a loss dominated by
exposures pulls the model the other way.

\subsubsection{Target selection and expansion.}
For each user sample we traverse all of its logged targets and expand them into
independent training instances, each paired with its own $I_b$. We drop targets
that carry only an exposure, with no click, cart, or order, and keep the instances
whose behavior is discriminative. Every surviving instance is a pair of a
behavior-labeled target and its matching instruction, so the supervision and the
instruction agree by construction. In the listwise daily setting, we pack the
$N$ surviving targets of a user context into a single sequence and
decode them together, where each target is one item written as its SID triple
$(s^1,s^2,s^3)$. Flattened over the three codebook levels, this gives the
SID-token sequence $Y=(y_1,\dots,y_L)$ of Section~\ref{sec:preliminaries} with
$L=3N$ decoding steps. The decoder thus produces a list of $N$ items in one pass
rather than a single item, and this listwise sequence is the same form the
verifiable reward and distillation operate on in post-training.

\subsubsection{Behavior-weighted objective.}
We weight the loss by behavior value
$w_b$ to each interaction type, giving
the pre-training objective
\begin{equation}
  \mathcal{L}_{\text{pre}}
  = \frac{1}{|\mathcal{T}|}\sum_{(y_t,b)\in\mathcal{T}}
    w_b\,\mathrm{CE}\big(p_\theta(y_t \mid X, I_s, I_r, I_b),\, y_t\big).
\end{equation}

Here $\mathcal{T}$ is the set of (SID token, behavior) pairs over all expanded
training instances in the batch: each target token $y_t$ inherits the behavior
label $b$ of the item it belongs to, and $w_b$ is that behavior's value weight.
The weight ordering matches the instruction priority and, later, the teacher's
behavior threshold, so a single behavior-value scale runs through the whole
pipeline.

\subsection{Post-training: Verifiable Rewards and Privileged Distillation}
We propose EA-TOSD, a post-training framework that couples
verifiable best-trajectory optimization with entropy-aware dual privileged
self-distillation. EA-TOSD optimizes the student trajectory 
by three components: selecting the optimal generated trajectory through a
verifiable reward, constructing a privileged teacher augmented with future
behavioral information, and performing entropy-aware dual self-distillation that
adapts supervision to the teacher's uncertainty.

\subsubsection{Verifiable rewards for trajectory optimization.}
In recommendation, reward signals~\cite{tulu3,deepseekr1} usually come from
a trained ranking model that scores each candidate. We instead use the hit rate of the generated SID as a verifiable reward. Let $X$ denote a user context sampled from the dataset
$\mathcal{D}$, and let $Y^{\star}=(y^{\star}_1,\ldots,y^{\star}_L)$ denote the
corresponding ground-truth SID token sequence, with $L=3N$ steps over the $N$
listwise targets. Given $X$, the current student policy generates a group of $G$
candidate sequences $\{\hat{Y}_i\}_{i=1}^{G}$, where $\hat{Y}_i=(\hat{y}_{i,1},\ldots,\hat{y}_{i,L})$ and
$\{\hat{Y}_i\}_{i=1}^{G}\sim\pi_S(\cdot\mid X)$. Here $\hat{y}_{i,t}$ is the SID token
generated at step $t$ of the $i$-th sampled sequence, and $\hat{y}_{i,<t}$ its
preceding prefix, so the student predicts the next token according to
$\pi_S(\cdot\mid X,\hat{y}_{i,<t})$. For a candidate sequence $i$, we
read off the per-step hits
\begin{equation}
  h_{i,t} = \mathbb{I}\!\left[ \hat{y}_{i,t} = y^\star_t\right], \qquad t=1,\dots,L,
\end{equation}
and aggregate them into a scalar  using a geometric weighting scheme. Earlier
decoding steps correspond to the coarser codebook levels that fix an item's broad
category.
\begin{equation}
  \omega_t = \frac{\gamma^{\,t-1}}{\sum_{j=1}^{L}\gamma^{\,j-1}},
  \qquad
  R_i = \sum_{t=1}^{L}\omega_t\, h_{i,t} \in [0,1],
  \label{eq:adv}
\end{equation}
where $\gamma$ is a geometric decay factor, $\omega_t$ is the weight of step $t$ and $R_i$ is the verifiable reward value of generated sequence $i$. For each context
$X$ sampled from $\mathcal{D}$, we draw $G$ candidate trajectories from the
current student policy and let $k$ denote the trajectory with the highest
verifiable score,
$R_{k}=\max_{i\in\{1,\ldots,G\}}R_i$. We then reinforce all tokens in the selected trajectory using its
verifiable score as a trajectory-level weight:
\begin{equation}
\label{eq:vbto_loss}
\begin{aligned}
\mathcal{L}_{\mathrm{VR}}
=
\mathbb{E}_{X\sim\mathcal D}
\Bigg[
-R_{k}
\sum_{t=1}^{L}
\log\pi_S\!\left(
\hat{y}_{k,t}\mid X, \hat{y}_{k,<t}
\right)
\Bigg].
\end{aligned}
\end{equation}

The expectation implicitly includes the randomness from sampling the candidate
group and selecting its highest-scoring trajectory. In the subsequent distillation Section~\ref{privileged teacher} and Section~\ref{sd}, we perform the subsequent distillation on the selected trajectory $k$.
\subsubsection{Privileged teacher structure.}
\label{privileged teacher}
Existing OPD methods~\cite{opd} rely on a separate, stronger teacher.
OPSD~\cite{opsd} lets a student model learn from its own generated trajectories under richer and more informative supervision, providing an effective post-training paradigm for LLMs. In industrial recommendation, a natural extension is to
equip the teacher with privileged inputs. We give the teacher the user's future
interactions as privileged information, aligning the model with user
preferences through behavioral signals. As Figure~\ref{fig:frame-overview}
shows, we take the items the user actually interacts with in the future,
including clicks, add-to-cart actions, and completed purchases, and prepend them
to the decoder input as a privileged decoding prefix $F$: a teacher-only block
built from the SIDs of at most $m$ near-future targets, with construction
details in Appendix~\ref{app:teacher}. The teacher shares the
student's backbone and differs only by this extra prefix, giving a student
policy $\pi_S$ and a teacher policy $\pi_T$. At each position of the
student-generated sequence, the privileged teacher produces a more confident
predictive distribution, so its distillation signals guide the student toward higher quality sequences.

\subsubsection{Entropy-aware dual self-distillation from a privileged teacher.}
\label{sd}
To complement the sparse hit-based signals, we use the privileged teacher to
provide dense token-level distillation supervision. Given the teacher policy
$\pi_T$ and the student policy $\pi_S$, the privilege advantage of predicted position $t$ is
\begin{equation}
\mathcal{A}_t
=
\log \pi_T\!\left(
\hat{y}_t \mid X, F, \hat{y}_{<t}
\right)
-
\log \pi_S\!\left(
\hat{y}_t \mid X, \hat{y}_{<t}
\right),
\end{equation}
where $\hat{y}_t$ is the SID token at step $t$ of the selected trajectory, and $F$ is
the training-only privileged prefix that encodes the user's near-future targets
(Appendix~\ref{app:teacher}). The term $\mathcal{A}_t$ transfers the teacher's future-informed preference signal to the student policy conditioned only on observable history. However, the privileged teacher does not provide beneficial distillation
signals at every prediction position. For each student generated prefix, we
compute the privileged teacher's predictive entropy at the next-token position
and use an entropy gate to decide whether the corresponding token-level
distillation loss is active. $H_{t}^T$ denotes the token-level entropy of the
teacher at step $t$:
\begin{equation}
    H_{t}^{T}=-\sum_{v\in\mathcal{V}}\pi_{T}(v\mid X, F, \hat{y}_{<t})\,\log\pi_{T}(v\mid X, F, \hat{y}_{<t}),
\end{equation}
where $v$ is a candidate token in the vocabulary $\mathcal{V}$. Low-entropy
positions indicate a clear token-level distillation direction, while
high-entropy positions capture multiple plausible future preferences. We
therefore use a binary low-entropy gate $g_t^l=\mathbb{I}\!\left({H}_{t}^T<\tau_l\right)$ and a high-entropy gate $g_t^{\mathrm{h}}
=
\mathbb{I}\!\left(
H_{t}^T>\tau_h
\right)$ to selectively distill preference information at each generation
position. The normalized weights $\widetilde g_t^l$ and $\widetilde g_t^h$ then
average the distillation loss over the selected tokens:
\begin{equation}
\widetilde g_t^l
=
\frac{g_t^l}{
\sum_{j=1}^{L} g_j^l+\epsilon
}
,\qquad
\widetilde g_t^h
=
\frac{g_t^h}{
\sum_{j=1}^{L} g_j^h+\epsilon
}.
\label{gt}
\end{equation}

Under the low-entropy gate, the privileged teacher supplies reliable
distillation signals with a clear user-preference direction, which realizes OPSD
at industrial recommendation scale. The resulting entropy-aware
self-distillation loss from the privileged teacher is
\begin{equation}
\begin{aligned}
\mathcal{L}_{\mathrm{SD}}
=\mathbb{E}_{
X\sim\mathcal D
}
\left[
-\sum_{t=1}^{L}
\widetilde g_t^l\,
\operatorname{sg}\!\left(\mathcal{A}_t\right)\,
\log\pi_S\!\left(
\hat{y}_t\mid X, \hat{y}_{<t}
\right)
\right],
\end{aligned}
\end{equation}
where $\operatorname{sg}(\cdot)$ is the stop-gradient operator. 
To preserve multiple plausible future user preferences and keep the student from prematurely collapsing onto a single mode, we further apply a forward KL divergence at the
high-entropy positions $(H_{t}^T>\tau_h)$, which transfers the teacher's uncertainty and global structure~\cite{eopd}. The
high-entropy FKL objective is
\begin{equation}
\label{eq:high_entropy_fkl}
\begin{aligned}
\mathcal{L}_{\mathrm{FKL}}
=
\mathbb{E}_{
X\sim\mathcal D
}
\left[
\sum_{t=1}^{L}
\widetilde g_t^{\mathrm{h}}\,
D_{\mathrm{KL}}\!\left(
\pi_T\!\left(
\cdot \mid X, F, \hat{y}_{<t}
\right)
\,\middle\|\,
\pi_S\!\left(
\cdot \mid X, \hat{y}_{<t}
\right)
\right)
\right].
\end{aligned}
\end{equation}

\begin{table*}[tbhp]
\centering
\small
\setlength{\tabcolsep}{5pt}
\caption{Main comparison under one backbone. Proxy-RM replaces our verifiable
reward with an external ranking model. \textbf{Bold} marks the best result in
each column; small \textcolor{teal}{green} numbers give OxygenREC-v2's gain over
the OxygenREC-v1 (PT-only) base.}
\label{tab:main}
\newcommand{\up}[1]{{\scriptsize\textcolor{teal}{(+#1)}}}
\begin{tabular}{@{}l cccc ccc@{}}
\toprule
& \multicolumn{4}{c}{\textbf{Retrieval}} & \multicolumn{3}{c}{\textbf{Ranking}} \\
\cmidrule(lr){2-5}\cmidrule(lr){6-8}
\textbf{Method} & HR@1 & HR@128 & HR@512  & Recall@512 & NDCG@512 & MRR@512 & GAUC \\
\midrule
OxygenREC-v1 (PT-only)  & 4.62\% & 33.08\% & 43.24\% & 34.95\% & 0.6211 & 0.5143 & 0.6207 \\
Proxy-RM      & 5.54\% & \textbf{33.89\%} & 43.09\% & 36.24\% & 0.6246 & 0.5343 & 0.6195 \\
\textbf{OxygenREC-v2} & \textbf{5.64\%}\up{1.02\%} & 33.76\%\up{0.68\%} & \textbf{44.14\%}\up{0.90\%} & \textbf{36.39\%}\up{1.44\%} & \textbf{0.6254}\up{0.0043} & \textbf{0.5369}\up{0.0226} & \textbf{0.6218}\up{0.0011} \\
\bottomrule
\end{tabular}
\end{table*}

\subsubsection{Training objective.}
Post-training optimizes three terms in parallel as a weighted sum: the
verifiable-reward objective $\mathcal{L}_{\text{VR}}$, the low-entropy privileged
distillation $\mathcal{L}_{\text{SD}}$, and the high-entropy forward-KL term
$\mathcal{L}_{\text{FKL}}$, with manually set weights $\lambda,
\beta, \zeta$.
\begin{equation}
  \mathcal{L}_{\text{EA-TOSD}} = \lambda\,\mathcal{L}_{\text{VR}}
              + \beta\,\mathcal{L}_{\text{SD}}
              + \zeta\,\mathcal{L}_{\text{FKL}}.
  \qquad
  \label{eq:objective}
\end{equation}

\begin{table*}[t]
\centering
\small
\setlength{\tabcolsep}{5pt}
\caption{Pre-training ablation, one component at a time.}
\label{tab:preablation}
\begin{tabular}{@{}l cccc ccc@{}}
\toprule
& \multicolumn{4}{c}{\textbf{Retrieval}} & \multicolumn{3}{c}{\textbf{Ranking}} \\
\cmidrule(lr){2-5}\cmidrule(lr){6-8}
\textbf{Ablation} & HR@1 & HR@128 & HR@512 & Recall@512 & NDCG@512 & MRR@512 & GAUC \\
\midrule
(a) OxygenREC-v1 (PT-only, w/o $I_b$)  & 4.62\% & 33.08\% & 43.24\% & 34.95\% & 0.6211 & 0.5143 & 0.6207 \\
(b) $+\,I_b$                  & 4.78\% & 33.47\% & 43.55\% & 35.11\% & 0.6223 & \textbf{0.5166} & \textbf{0.6218} \\
(c) $+\,w_b$
                              & \textbf{4.91\%} & \textbf{33.63\%} & \textbf{43.82\%} & \textbf{35.25\%} & \textbf{0.6224} & \textbf{0.5166} & \textbf{0.6218} \\
\bottomrule
\end{tabular}
\end{table*}

The supervised fine-tuning next-token loss (SFT) runs alongside the reinforcement update rather than
alternating with it. It anchors the policy to the pre-training distribution and
adds dense supervision next to the sparse hit, playing the role of a
pre-training-mix (PTX) regularizer. With the external ranking model removed, the
reward reduces to one verifiable signal, and the supervised and distillation
terms act as complementary regularizers.

\section{Experimental Setup}
\label{sec:setup}

\paragraph{Data.}
We train and evaluate on proprietary interaction logs from JD.com, one of the largest industrial
e-commerce platforms in the world. Each example pairs a user context with the items the user
engaged with, labeled by behavior type (exposure, click, add-to-cart, order),
and we follow the listwise daily setting used in pre-training, where a user
context is matched with a group of engaged items decoded as one packed sequence.
Training covers one month of logs and evaluation uses the following day. Full
dataset statistics and the item tokenizer are given in
Appendix~\ref{app:setup}.

\paragraph{Metrics.}
We measure two abilities. For \emph{retrieval}, we report HR@$K$ (Hit Rate) and
Recall@$K$ at $K\in\{1,128,512\}$, which capture whether the generated candidates
recover the engaged items. For \emph{ranking}, we report NDCG@512 (Normalized Discounted Cumulative Gain) and MRR@512 (Mean Reciprocal Rank)
over the decoded list, together with GAUC (Group Area Under the Curve): the model
scores the logged candidate list, and GAUC is computed per user and averaged over
the test set, both overall and separately for click, cart, and order. Higher is
better for all metrics.

\paragraph{Compared methods.}
We compare three variants under the same backbone and decoding format.
\textbf{OxygenREC-v1 (PT-only)} is the pre-trained checkpoint without post-training by following the same setting in ~\cite{oxygenrec}.
\textbf{Proxy-RM} follows the external discriminator pattern of the prior
industrial system~\cite{zhou2025onerectechnicalreport,zhang2026onemall,oxygenrec} with a unified ranking model
used as a proxy reward by finetuning OxygenREC-v1 (PT-only). \textbf{OxygenREC-v2} is our proposed method. All
variants share one backbone and add no inference-time module. All offline studies
use a 3B encoder-decoder backbone. Backbone size,
optimizer, and default hyperparameters are listed in
Appendix~\ref{app:setup}.

\paragraph{Online A/B protocol.}
We deploy OxygenREC-v2 as a 3B-parameter, 1B-activated (3B-A1B) MoE on JD.com's
industrial e-commerce platform and run bucketed A/B tests on six production
surfaces, each compared to models then in production. Buckets receive
$5$--$20\%$ of surface traffic and run for five to eight days; we report relative
lift on UCTR, UCTCVR and GMV,
with significance from
the platform's standard two-sided test. Surfaces are grouped by the user-lifecycle
stages of OxygenREC-v1~\cite{oxygenrec}, spanning the homepage floors, channel
feeds, checkout paths, and product-detail pages.

\section{Experimental Results}
\label{sec:result}

\subsection{Main Results}
Table~\ref{tab:main} compares three models on the same backbone: the pre-trained
OxygenREC-v1 (PT-only) checkpoint, the proxy-reward baseline that follows the external
discriminator pattern (Proxy-RM), and our full model OxygenREC-v2. We evaluate
two abilities: \emph{retrieval}, whether the generated candidates recover the
engaged items (HR@$K$ and Recall@$K$), and \emph{ranking}, whether items that
received a behavior are ranked higher under the matching instruction (NDCG@512,
MRR@512, and GAUC).

OxygenREC-v2 wins the retrieval metrics that decide the candidate set. It reaches
44.14\% HR@512 and 36.39\% Recall@512, ahead of OxygenREC-v1 by 0.90 and 1.44
points and ahead of Proxy-RM by 1.05 and 0.15. The gap widens at the top of the
list: HR@1 rises from 4.62\% to 5.64\%, a 1.02-point gain over v1 that lifts the
hardest retrieval metric the most. On ranking, OxygenREC-v2 also leads on all
three metrics, with NDCG@512 0.6254, MRR@512 0.5369, and GAUC 0.6218, improving
over v1 by 0.0043, 0.0226, and 0.0011.

Proxy-RM shows why an external reward is not enough. It improves ranking over v1
(0.6246 NDCG, 0.5343 MRR) but stalls on HR@512 (43.09\%, below v1's 43.24\%): the
proxy reward reorders candidates without pulling behavior-relevant items into the
set. OxygenREC-v2 avoids this by conditioning generation on the target behavior,
so the retrieval and ranking gains come together. 

\noindent\textbf{Takeaway.} Internalizing the behavior signal, rather than
attaching it as an external reward, improves the candidate set itself: OxygenREC-v2
leads on six of seven metrics and gains most on HR@1 and Recall@512, the metrics
that reward retrieving the right items rather than reordering a fixed list.

\subsection{Ablation Study}
\subsubsection{Pre-training ablation.}
Table~\ref{tab:preablation} isolates the pre-training components by adding them
one at a time on the same backbone: (a) the Base generator without the
behavior instruction; (b) $+\,I_b$, adding the behavior instruction; and (c)
$+\,w_b$, further adding the behavior-weighted loss, which is our full
pre-training. On the same retrieval and ranking metrics as the main
comparison, both components improve retrieval monotonically: HR@512 rises from
$43.24\%$ (Base) to $43.55\%$ with $I_b$ and to $43.82\%$ with the behavior-weighted
loss, and Recall@512 follows the same order ($34.95\%\!\to\!35.11\%\!\to\!35.25\%$).
NDCG@512 is highest for the full configuration, while MRR@512 and GAUC saturate
once $I_b$ is added, tying at $0.5166$ and $0.6218$ for (b) and (c). Each
component contributes a small but consistent gain, and the two are complementary
rather than redundant.

\subsubsection{Privileged teacher.}
\begin{table}[htbp]
\centering
\caption{Ablation on the privileged information used to augment the teacher.}
\label{tab:privileged_information_ablation}
\resizebox{0.9\columnwidth}{!}{
\begin{tabular}{lccc}
\toprule
\textbf{Method}
& \textbf{HR@1}
& \textbf{HR@512}
& \textbf{Recall@512} \\
\midrule

Proxy-RM
& 5.54\%
& 43.09\%
& 36.24\% \\

Privileged Teacher (Expo)
& 5.62\%
& 43.80\%
& 36.19\% \\

Privileged Teacher (Click)
& \textbf{5.64\%}
& \textbf{44.14\%}
& \textbf{36.39\%} \\

\bottomrule
\end{tabular}
}
\end{table}
The selection of privileged information for the teacher is not arbitrary. Table~\ref{tab:privileged_information_ablation} compares different choices of future interaction behaviors used to augment the teacher. In the Expo setting, the privileged information includes all future interacted items, covering exposure-only, exposure-and-click, click-and-add-to-cart, and add-to-cart-and-purchase behaviors. In the Click setting, exposure-only items are excluded, and the privileged information consists of items associated with exposure-and-click, click-and-add-to-cart, and add-to-cart-and-purchase behaviors.

The results show that the Click setting achieves the best overall performance, reaching 5.64\% HR@1, 44.14\% HR@512, and 36.39\% Recall@512. In contrast, including exposure-only items slightly reduces HR@512 and Recall@512 to 43.80\% and 36.19\%, respectively. This suggests that exposure only items add noise; clicked items associated with deeper interactions give cleaner privileged signals.

\subsubsection{Training objectives.}
To investigate the contribution of each training objective, we progressively
remove individual loss components from the complete post-training objective
($\lambda,\beta,\zeta$ in Eq.~\ref{eq:objective}; default $\lambda{=}0.1$,
$\beta{=}\zeta{=}0.01$). w/o FKL keeps entropy-aware self-distillation
from the privileged teacher but drops the high-entropy FKL term ($\zeta{=}0$);
w/o FKL\&SD enables only the verifiable best-trajectory objective
($\beta{=}\zeta{=}0$); SFT only is the supervised baseline without
post-training ($\lambda{=}\beta{=}\zeta{=}0$); and EA-TOSD activates all
components. Table~\ref{tab:objective_ablation} lists the weights alongside the
metrics.

\begin{table}[htbp]
\centering
\caption{Ablation on post-training objectives, with the loss weights of
Eq.~\ref{eq:objective}.}
\label{tab:objective_ablation}
\resizebox{0.8\columnwidth}{!}{
\begin{tabular}{lcccccc}
\toprule
\textbf{Method} &
\textbf{HR@1} &
\textbf{HR@512} &
\textbf{Recall@512} \\
\midrule
\textbf{EA-TOSD}

& \textbf{5.64\%}
& \textbf{44.14\%}
& \textbf{36.39\%} \\
w/o FKL

& 5.59\%
& 43.80\%
& 36.21\% \\
w/o FKL\&SD

& \textbf{5.64\%}
& 43.65\%
& 36.14\% \\
SFT only

& 5.52\%
& 43.65\%
& 36.20\% \\
\bottomrule
\end{tabular}
}
\end{table}
Table~\ref{tab:objective_ablation} presents the ablation results for different post-training objectives. Starting from the complete EA-TOSD framework, removing FKL degrades HR@512 from 44.14\% to 43.80\% and Recall@512 from 36.39\% to 36.21\%, indicating that entropy-aware FKL provides complementary supervision for uncertain prediction positions. Further removing privileged self-distillation reduces HR@512 to 43.65\% and Recall@512 to 36.14\%, demonstrating the effectiveness of dense token-level supervision from the privileged teacher. Finally, further removing VR leads to a decrease in HR@1 from 5.64\% to 5.52\%, confirming that the proposed trajectory optimization and entropy-aware self-distillation jointly contribute to the performance gains.

\subsubsection{Entropy-based token routing.}Since the privileged teacher does not provide equally reliable supervision at every prediction position, we investigate how the coverage of entropy-based routing affects the final recommendation performance. Specifically, we vary the entropy thresholds $(\tau_l,\tau_h)$  to control the proportions of prediction positions routed to SD and FKL, denoted by $\rho_L$ and $\rho_H$.
As shown in Table~\ref{tab:entropy_routing_ablation}, applying distillation to all prediction positions (100\%/100\%) does not achieve the best performance. Reducing the routing coverage to 20\% for both SD and FKL improves HR@1 from 5.42\% to 5.64\%, HR@512 from 43.87\% to 44.14\%, and Recall@512 from 36.27\% to 36.39\%. Increasing the coverage to 40\% also underperforms the 20\% setting. These results demonstrate that only a subset of prediction positions provides reliable privileged supervision, validating the effectiveness of entropy-based token routing over indiscriminate full-token distillation.
\begin{table}[hbtp]
\centering
\caption{Ablation on entropy-based token routing.}
\label{tab:entropy_routing_ablation}
\resizebox{1.0\columnwidth}{!}{
\begin{tabular}{lccc}
\toprule
\textbf{Distillation Token Ratio}
& \textbf{HR@1}
& \textbf{HR@512}
& \textbf{Recall@512} \\
\midrule
$\rho_{\mathrm{L}}=\rho_{\mathrm{H}}=100\%$
& 5.42\%
& 43.87\%
& 36.27\% \\

$\rho_{\mathrm{L}}=\rho_{\mathrm{H}}=40\%$
& 5.44\%
& 43.80\%
& 36.21\% \\

$\rho_{\mathrm{L}}=\rho_{\mathrm{H}}=20\%$
& \textbf{5.64\%}
& \textbf{44.14\%}
& \textbf{36.39\%} \\
\bottomrule
\end{tabular}
}
\end{table}

\subsection{Online A/B Testing}
\label{sec:online-ab}
We evaluate OxygenREC-v2 online with bucketed A/B tests on six production
surfaces\footnote{Scenario 1: Budget \& Free Shipping homepage floor; Scenario 2: Billion Subsidy
channel feed; Scenario 3: Budget \& Free Shipping channel feed; Scenario 4:
Seckill feed; Scenario 5: Add-to-Cart Overlay; Scenario 6: RecForU.}, each compared
against the model then serving it. Table~\ref{tab:online} reports the lift on
UCTR, UCTCVR and GMV.

The pre-training upgrade (listwise decoding with the behavior instruction) drives
conversion on the feed and floor surfaces: UCTCVR rises by $1.6$--$4.4\%$ and is
significant on every one of them, while UCTR stays essentially flat, so the upgrade
changes \emph{which} items convert rather than inflating clicks. GMV moves in the
same direction but is noisier, as expected for a long-tail-dominated metric,
reaching significance on the homepage floor ($+21.2\%$) and staying directional on
the feeds. The add-to-cart overlay (Scenario~5) is the lone exception: its GMV is
slightly negative but not significant ($-2.3\%$, $p{=}0.23$), within day-to-day
fluctuation rather than a real regression. The post-training upgrade (verifiable
reward with privileged distillation), evaluated on the product-detail surface
(Scenario~6) under a unified all-scene policy, lifts GMV by $4.3\%$ ($p{=}0.048$).
Across surfaces the gains are consistent in sign and, on the metrics that decide
revenue, reach significance despite the short measurement windows.

\begin{table}[t]
\centering
\small
\setlength{\tabcolsep}{5pt}
\caption{Online A/B results across deployment surfaces, grouped by
user-lifecycle stage (relative lift over the production model; $^\ast$ significant
at $p<0.05$).}
\label{tab:online}
\resizebox{\columnwidth}{!}{%
\begin{tabular}{@{}ll ccc@{}}
\toprule
\textbf{Stage} & \textbf{Surface} & \textbf{UCTR} & \textbf{UCTCVR} & \textbf{GMV} \\
\midrule
Homepage Floor & Scenario 1 & $+0.09\%$      & $+3.62\%^\ast$ & $+21.21\%^\ast$ \\
\midrule
\multirow{3}{*}{Channel Feeds}
               & Scenario 2 & $+0.08\%$      & $+1.61\%^\ast$ & $+2.80\%$  \\
               & Scenario 3 & $+1.39\%^\ast$ & $+4.44\%^\ast$ & $+6.86\%$  \\
               & Scenario 4 & $+0.01\%$      & $+2.90\%^\ast$ & $+1.09\%$  \\
\midrule
\multirow{2}{*}{Checkout Path} 
               & Scenario 5 & $-0.34\%$      & $+2.15\%^\ast$      & $-2.34\%$  \\
               & Scenario 6 & $-0.03\%$      & $+0.56\%$      & $+4.25\%^\ast$ \\
\bottomrule
\end{tabular}%
}
\end{table}

\section{Analysis: How the Behavior Instruction Reshapes Generation}
\label{sec:discuss-behavior}

\begin{figure}[t]
    \centering
    \includegraphics[width=0.8\columnwidth]{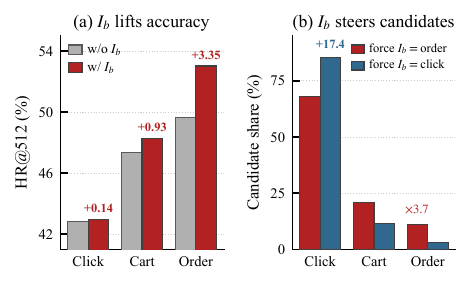}
    \caption{Behavior conditioning reshapes generation. \textbf{(a)} The behavior
instruction $I_b$ improves per-behavior HR@512, with the largest gain on the
sparse, high-value order behavior. \textbf{(b)} Forcing $I_b$ to a fixed behavior
shifts the generated candidate set toward that behavior, showing that $I_b$
changes what is generated rather than only how candidates are scored.}
    \Description{Two side-by-side bar charts. The left panel shows per-behavior
    HR@512 for click, cart, and order with and without the behavior instruction,
    with the largest improvement on order. The right panel shows the candidate-set
    behavior composition when the behavior instruction is forced to order versus
    click, with the order share rising under a forced order intent and the click
    share rising under a forced click intent.}
\label{fig:behavior}
\end{figure}

Does the behavior instruction $I_b$ genuinely steer generation or does the model
learn to ignore it as a passive label? We probe the pre-trained model directly:
for a fixed set of user contexts, we vary only the injected $I_b$ and inspect both
retrieval quality and the composition of the generated candidates.
Figure~\ref{fig:behavior} summarizes the two findings.

\paragraph{Behavior conditioning improves the sparse, high-value behaviors most.}
Figure~\ref{fig:behavior}(a) reports the per-behavior HR@512 with and without the
behavior instruction, conditioning each case on its true target behavior. Adding
$I_b$ lifts every behavior, but the gain concentrates on the order behavior (the
sparsest yet highest-value signal), which improves far more than click. The
behavior signal is most useful precisely where behavior-agnostic generation has
the least to go on, directing modeling capacity toward the conversions the
platform cares about.

\paragraph{Forcing the instruction steers the candidate set.}
Figure~\ref{fig:behavior}(b) tests whether $I_b$ controls generation rather than
just annotating it. We override the instruction at inference, decoding every
context under a single fixed behavior, and measure the behavior composition of
the generated Top-$K$ candidates using a leakage-free item--behavior prior. The
candidate mix shifts markedly toward the forced behavior: forcing order pulls the
set toward order-associated items, while forcing click pulls it toward
click-associated ones. This also holds under the true-target metric: forcing order
holds the order subset at $53.0$ HR@512 while dropping click to $34.5$, and forcing
click holds click but loses nearly five points on order (App.~\ref{app:instruct_steering}).
The same user and model produce visibly different candidate sets under different
behavior instructions.

Output-side alternatives cannot provide this property. A behavior classifier or an
autoregressive behavior suffix only reorders a candidate set that was already
generated behavior-agnostically, changing how candidates are \emph{scored}, not
\emph{which} are produced. Placing the signal on the input side, as $I_b$ does,
moves the intervention upstream into generation itself, so the candidate
distribution becomes behavior-aware by construction, which is exactly what the
downstream verifiable reward then reinforces.

\begin{figure}[t]
    \centering
    \includegraphics[width=0.8\columnwidth]{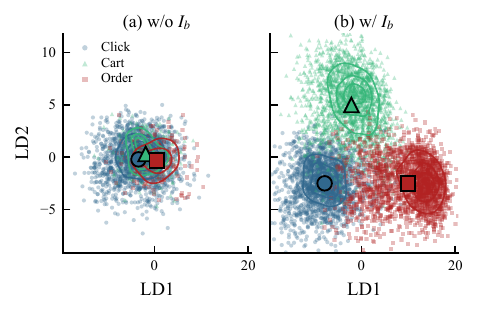}
    \caption{First-step decoder hidden states projected onto the first two linear
discriminant axes (LD1, LD2) of a behavior-supervised LDA, colored by behavior.
\textbf{(a) w/o $I_b$:} the three behaviors collapse into one
cluster. \textbf{(b) w/ $I_b$:} they separate, showing that the instruction
conditions the generation state before the first token.}
    \Description{Two side-by-side scatter plots of decoder hidden states projected
    to two dimensions. In the left panel the click, cart, and order points overlap
    in a single cluster; in the right panel they form three separated clusters.}
\label{fig:separation}
\end{figure}

\paragraph{The behavior signal separates the internal generation state.}
The two effects above are measured on the model's outputs; we now look inside the
decoder. For each context, we take the decoder hidden state at the first decoding
step, after the prefix $[\textsc{bos}, I_s, I_r, I_b]$ is consumed but before any SID
token is emitted, which is where $I_b$ can steer generation. Projecting these
states onto the first two linear discriminant axes of a behavior-supervised LDA
(Figure~\ref{fig:separation}), the
behavior-agnostic model collapses the three behaviors into one indistinguishable
cloud, whereas our model resolves them into three separated clusters. The
instruction thus reorganizes the generation state by target behavior before the
first token is produced, the mechanism underlying the candidate-set shift reported
above.

\section{Conclusion}
\label{sec:conclusion}
We asked whether the discriminative behavior signal in generative recommendation,
the clicks, cart-adds, and orders, can be built into the generator's training
objective rather than added afterward by a separate reward model. We presented
OxygenREC-v2, which internalizes discrimination into generative recommendation
(IDGR): it conditions generation on behavior from the first decoding step and
replaces the proxy reward with a verifiable behavior-match reward and privileged
distillation, all on one shared backbone. The behavior signal steers what the
model generates, not just how outputs are scored, with the largest gains on order,
the sparsest and highest-value behavior. On JD.com's e-commerce platform,
OxygenREC-v2 improves offline retrieval and ranking over the external-reward
baseline, and, deployed as a 3B-A1B MoE, raises online UCTCVR and
GMV in bucketed A/B tests. These results position the behavior signal as a
first-class training objective for generative recommenders, not a post-hoc
scoring signal. Future work extends beyond the click/cart/order hierarchy and
reduces the training cost of the privileged teacher.

\printbibliography

\clearpage
\appendix

\section{Notation}
\label{app:notation}

Table~\ref{tab:notation} collects the symbols used in the preliminaries and
methodology. Two granularities recur throughout: the \emph{item} level, where a
recommendation is a list of $N$ items, and the \emph{token} level, where the
same list is a flattened sequence of $L=3N$ SID tokens indexed by the decode
step $t$.

\section{Behavior Instruction Construction}
\label{app:instruct}

This section gives the implementation details of the behavior-aware
instruction $I_b$ introduced in the methodology. The main text states the
construction compactly as $I_b=\psi(\mathrm{Emb}(b))$ and the decoder prefix
as $[\textsc{bos}, I_s, I_r, I_b]$; here we specify the token mapping, the
projection network $\psi$, and the prefix layout, then evaluate where $I_b$
enters the prefix (Appendix~\ref{app:instruct_position}) and whether it actually
steers generation (Appendix~\ref{app:instruct_steering}).

\subsection{Behavior Token and Embedding Lookup}
\label{app:instruct_token}

The target behavior $b\in\{\mathrm{click},\mathrm{cart},\mathrm{order}\}$ is
first mapped to a reserved vocabulary token. We assign the three behaviors to
consecutive identifiers placed above the SID range, so that the
behavior tokens never collide with item codes,
\begin{equation}
\begin{aligned}
    &\mathrm{tok}(b) = \tau_0 + \rho(b), \\
  &\rho(\mathrm{click})\!=\!1,\;
  \rho(\mathrm{cart})\!=\!2,\;
  \rho(\mathrm{order})\!=\!3,
\end{aligned}
\end{equation}

where $\tau_0$ is a fixed offset that reserves a contiguous block of behavior
tokens. The token is then embedded through the decoder's \emph{shared}
embedding table $\mathrm{Emb}_{\mathrm{dec}}(\cdot)\in\mathbb{R}^{d_{\text{model}}}$,
the same table used to embed SID tokens,
\begin{equation}
  e_b = \mathrm{Emb}_{\mathrm{dec}}\!\big(\mathrm{tok}(b)\big)
        \in\mathbb{R}^{d_{\text{model}}}.
\end{equation}
Reusing the decoder embedding table, rather than introducing a separate
behavior embedding, keeps the behavior signal in the same representation space
as the tokens the decoder generates and adds no new lookup parameters.

\subsection{Projection Network $\psi$}
\label{app:instruct_proj}

The behavior embedding $e_b$ is passed through a lightweight two-layer
feed-forward projection $\psi$ that maps it to the instruction
$I_b\in\mathbb{R}^{d_{\text{model}}}$,
\begin{equation}
  I_b = \psi(e_b)
      = W_2\,\big(\mathrm{drop}\big(\sigma(W_1 e_b)\big)\big),
\end{equation}
with $W_1\in\mathbb{R}^{d_{\text{ff}}\times d_{\text{model}}}$,
$W_2\in\mathbb{R}^{d_{\text{model}}\times d_{\text{ff}}}$, a LeakyReLU
nonlinearity $\sigma$ (negative slope $0.01$), and dropout between the layers;
the hidden width matches the backbone feed-forward size $d_{\text{ff}}$. The
projection is initialized from the backbone initializer and trained jointly
with the model. An optional RMSNorm may precede the first layer when the
instruction adapter is normalized; it is disabled by default. The projection
$\psi$ is the only parameter added for the behavior instruction, and it is
shared across the three behaviors.

\subsection{Decoder Prefix Layout}
\label{app:instruct_prefix}

The behavior instruction is concatenated along the sequence dimension after
the scene and reasoning instructions $I_s, I_r$, and the BOS embedding is prepended, giving the
decoder prefix
\begin{equation}
  \big[\,\textsc{bos},\; I_s,\; I_r,\; I_b\,\big],
\end{equation}
followed by the SID tokens of the target. The decoder attends to this
prefix at every decoding step, so $I_b$ conditions the entire autoregressive
generation. The same layout is used at training and inference; the only
difference is the source of $b$, which is the logged target behavior during
training and the scene's business goal at serving time.

\subsection{Instruction Placement}
\label{app:instruct_position}

The relative order of the BOS embedding and the instruction block
$[I_s, I_b]$ is a configurable choice that we treat as an ablation axis. Let
$E=[I_s, I_b]$ denote the instruction embeddings. We consider three placements
that expose the decoder to the same content but differ in position:
\textbf{insert-right} (default), $[\textsc{bos}, E]$, the instruction follows the
BOS token; \textbf{insert-left}, $[E, \textsc{bos}]$, the instruction precedes
it; and \textbf{add-to-BOS}, $\textsc{bos}+E$, the single instruction embedding
is added to the BOS embedding rather than concatenated (applicable when the
instruction reduces to one token).

Table~\ref{tab:beh_placement} evaluates these placements against a no-instruction
baseline, each conditioned on the sample's actual behavior; all numbers are HR at
Top-$512$ in \% ($\uparrow$), on the same window and backbone as the main
pre-training experiments. Any placement of $I_b$ improves over the baseline, and
the gain concentrates on the order subset---the sparsest, highest-value
behavior---which rises from $49.67$ to $53.02$. The three placements are within
$0.3$ of one another on the overall rate, so the model is insensitive to where
$I_b$ sits relative to the BOS token; we adopt insert-right by default as it is
strongest on the order and click subsets.

\begin{table}[t]
\centering
\small
\setlength{\tabcolsep}{5pt}
\caption{Placement of the behavior instruction $I_b$, under each sample's actual
behavior. ``None'' is the BOS-only baseline; columns are HR (\%, $\uparrow$)
overall at Top-$1$/$512$ and the per-behavior breakdown at Top-$512$.}
\label{tab:beh_placement}
\begin{tabular}{@{}l cc ccc@{}}
\toprule
& \multicolumn{2}{c}{\textbf{Overall}} & \multicolumn{3}{c}{\textbf{Per-behavior @512}} \\
\cmidrule(lr){2-3}\cmidrule(lr){4-6}
\textbf{Placement} & @1 & @512 & ord & cart & clk \\
\midrule
None (BOS only)          & 4.85 & 43.60 & 49.67 & 47.36 & 42.85 \\
Insert-left              & 4.96 & 43.89 & 52.76 & 48.01 & 42.92 \\
Insert-right (default)   & \textbf{5.02} & 43.99 & \textbf{53.02} & \textbf{48.29} & 42.99 \\
Add-to-BOS               & 4.97 & \textbf{44.16} & 52.97 & 48.27 & \textbf{43.17} \\
\bottomrule
\end{tabular}
\end{table}

\subsection{Behavior Steering}
\label{app:instruct_steering}

The placement study conditions on each sample's true behavior. To test whether
$I_b$ \emph{controls} generation rather than merely correlating with it, we
override the instruction at inference: every sample is decoded under a single
fixed behavior, regardless of its true target, and scored against the true target
on each behavior subset. If $I_b$ steers the candidate set, forcing a behavior
should help the matching subset and hurt the others.
Table~\ref{tab:beh_steering} confirms this under the insert-right layout: the
best score on each subset lies on the diagonal, at or within noise of the
actual-behavior upper bound. Forcing order holds the order subset at $53.01$
(vs.\ $53.02$ actual) while dropping the click subset from $42.99$ to $34.54$;
forcing click holds the click subset at $42.99$ but loses nearly five points on
order ($53.02$ to $48.33$). The instruction does not just annotate the
target---it redirects generation toward the conditioned behavior, which is why
placing the behavior label on the input side outperforms the output-side suffix
variants of Appendix~\ref{app:suffix}.

\begin{table}[t]
\centering
\small
\setlength{\tabcolsep}{6pt}
\caption{Behavior steering under insert-right. Each row forces $I_b$ to a fixed
behavior; columns are HR@512 (\%, $\uparrow$) on the click, cart, and order
subsets. ``Actual'' is the true-behavior upper bound; the diagonal is in bold.}
\label{tab:beh_steering}
\begin{tabular}{@{}l ccc@{}}
\toprule
& \multicolumn{3}{c}{\textbf{Evaluation subset (HR@512)}} \\
\cmidrule(lr){2-4}
\textbf{Forced $I_b$} & clk & cart & ord \\
\midrule
Actual        & 42.99 & 48.29 & 53.02 \\
\midrule
Force clk     & \textbf{42.99} & 46.65 & 48.33 \\
Force cart    & 40.20 & \textbf{48.29} & 50.02 \\
Force ord     & 34.54 & 44.85 & \textbf{53.01} \\
\bottomrule
\end{tabular}
\end{table}

\section{Suffix-Based Behavior Modeling}
\label{app:suffix}

The behavior-aware instruction $I_b$ places the
behavior label on the \emph{input} side of the decoder, conditioning generation
on the target behavior. A natural alternative is to place the label on the
\emph{output} side and let the model predict it. We explore three such suffix
variants---Suffix AR, Suffix Head, and Suffix MultiTower---which share the same
backbone and SID tokenizer as our main model and differ only in how the behavior
label enters the training objective and inference procedure.

\subsection{Suffix AR: Behavior as an Autoregressive Token}
\label{app:suffix_ar}

Suffix AR extends the SID sequence with a fourth autoregressive step that decodes
the behavior type $b \in \{\mathrm{clk}, \mathrm{cart}, \mathrm{ord}\}$. Item
identity and behavior intent are jointly modeled by the same generator:
\begin{equation}
\begin{aligned}
  p_\theta\bigl(s^1, s^2, s^3, b \mid X, I_s\bigr)
  &= \prod_{t=1}^{3} p_\theta\bigl(s^t \mid s^{<t}, X, I_s\bigr) \\
  &\quad \cdot\, p_\theta\bigl(b \mid s^{1:3}, X, I_s\bigr).
\end{aligned}
\end{equation}
Training optimizes the standard weighted next-token loss over all four positions;
no auxiliary head is introduced. At inference, beam search produces hypotheses
of the form $(s^1, s^2, s^3, \hat{b})$, ranked by the joint log-probability.
Behavior is decoded \emph{after} the item, so it cannot
influence which items are proposed---only how they are ranked within the beam.
The candidate distribution itself remains behavior-agnostic.

\subsection{Suffix Head: Behavior as an Auxiliary Classifier}
\label{app:suffix_head}

Suffix Head decouples generation from behavior prediction. The decoder produces
the SID sequence as usual, and a lightweight three-way classification head
$\mathcal{F}_{\mathrm{CE}}$ is applied to the final decoder hidden state $h_b$
after the full SID has been consumed. The training objective combines item
generation with an auxiliary behavior loss:
\begin{equation}
  \mathcal{L}_{\mathrm{SuffixHead}}
  \;=\; \mathcal{L}_{\mathrm{NTP}} \;+\; \lambda \,\mathrm{CE}\bigl(\mathcal{F}_{\mathrm{CE}}(h_b),\, y_b\bigr),
\end{equation}
where $y_b$ is the ground-truth behavior label and $\lambda$ balances the two
terms.

\paragraph{Inference: generate-then-rerank.} Generation proceeds in two stages.
Stage~1 runs beam search over the SID space to produce $K$ candidates
$\{(s^1_k, s^2_k, s^3_k)\}_{k=1}^K$, free of any behavior conditioning. Stage~2
scores each candidate by feeding its SID back through the decoder and reading
off the behavior probabilities $\mathbf{p}^{(k)} \in \Delta^3$ from the head.
The candidates are then reranked by the component matching the deployment
target $b^*$:
\begin{equation}
  \mathrm{score}^{(k)} \;=\; \mathbf{p}^{(k)}_{b^*}.
\end{equation}
The reranking strategy can be changed at serving time without retraining---a
checkout-path scenario reranks by $P(\mathrm{ord})$, a homepage scenario by
$P(\mathrm{clk})$.
Behavior prediction is decoupled from item generation,
giving the reranking step full flexibility. As with Suffix AR, however, the
generated candidate set is itself behavior-agnostic; the head only reorders an
already-fixed beam.

\subsection{Suffix MultiTower: Independent Behavior Towers}
\label{app:suffix_multitower}

Suffix MultiTower treats the three behaviors as independent binary labels rather
than mutually exclusive classes. Three classification towers
$\mathcal{F}_{\mathrm{clk}}, \mathcal{F}_{\mathrm{cart}}, \mathcal{F}_{\mathrm{ord}}$
are attached to the same hidden state $h_b$, and the auxiliary loss is a sum of
binary cross-entropy terms:
\begin{equation}
  \mathcal{L}_{\mathrm{beh}}
  \;=\; \sum_{t \in \{\mathrm{clk}, \mathrm{cart}, \mathrm{ord}\}}
        \mathrm{BCE}\bigl(\mathcal{F}_t(h_b),\, y_t\bigr).
\end{equation}
This relaxes the single-label assumption of Suffix Head: a clicked item that
was also added to cart contributes positive supervision to both towers.
Inference follows the same generate-then-rerank pattern as Suffix Head, with
the per-tower sigmoid probabilities serving as reranking scores. When the
deployment context carries multiple co-occurring behavior targets, the
reranking score can be a weighted combination of tower outputs.

\subsection{Comparison and Empirical Results}
\label{app:suffix_comparison}

The three variants share the backbone, SID tokenizer, and one-month training
window of our main model and differ only in how the behavior label enters the
objective and inference. Each is cold-started and trained to convergence
($\approx\!294$k steps) and evaluated with beam size $512$, reporting
HR (hit rate) and Recall at Top-$1$ and Top-$512$, with the
per-behavior breakdown (order, cart, click) at Top-$512$. For Suffix~Head and
Suffix~MultiTower we report both the raw beam (\emph{beam}) and the reranked
output (\emph{+rerank}); Suffix~AR carries the behavior inside the beam and has
no separate reranking stage. Table~\ref{tab:suffix} reports the results.

\begin{table*}[t]
\centering
\small
\setlength{\tabcolsep}{5.5pt}
\caption{Suffix-based behavior-modeling variants (Appendix~\ref{app:suffix}),
in \% ($\uparrow$). HR and Recall at Top-$1$/$512$, with the per-behavior
breakdown at Top-$512$. \emph{beam} is the raw beam and \emph{+rerank} the
generate-then-rerank output; Top-$512$ columns are unchanged by reranking. Best
in bold.}
\label{tab:suffix}
\begin{tabular}{@{}ll ccccc ccccc@{}}
\toprule
& & \multicolumn{5}{c}{\textbf{HR}} & \multicolumn{5}{c}{\textbf{Recall}} \\
\cmidrule(lr){3-7}\cmidrule(lr){8-12}
& & @1 & \multicolumn{4}{c}{@512} & @1 & \multicolumn{4}{c}{@512} \\
\cmidrule(lr){4-7}\cmidrule(lr){9-12}
\textbf{Variant} & \textbf{Stage} & all & all & ord & cart & clk & all & all & ord & cart & clk \\
\midrule
Suffix AR              & beam    & 4.74 & 42.42 & 47.58 & 45.45 & 41.79 & 2.92 & 34.17 & 40.67 & 39.01 & 33.67 \\
\addlinespace[1pt]
\multirow{2}{*}{Suffix Head}
                       & beam    & 4.11 & \textbf{43.66} & \textbf{50.04} & 47.59 & \textbf{42.86} & 2.50 & \textbf{35.37} & 43.13 & 41.21 & \textbf{34.73} \\
                       & +rerank & \textbf{4.88} & \textbf{43.66} & \textbf{50.04} & 47.59 & \textbf{42.86} & \textbf{2.99} & \textbf{35.37} & 43.13 & 41.21 & \textbf{34.73} \\
\addlinespace[1pt]
\multirow{2}{*}{Suffix MultiTower}
                       & beam    & 4.44 & 43.65 & 50.01 & \textbf{47.62} & 42.84 & 2.70 & 35.32 & \textbf{43.19} & \textbf{41.28} & 34.68 \\
                       & +rerank & 4.87 & 43.65 & 50.01 & \textbf{47.62} & 42.84 & \textbf{2.99} & 35.32 & \textbf{43.19} & \textbf{41.28} & 34.68 \\
\bottomrule
\end{tabular}
\end{table*}

Suffix~AR is the weakest at Top-$512$ (HR $42.42$, Recall $34.17$), trailing
the other two by more than a point. Decoding the behavior as a fourth
autoregressive token spends a decoding step on a label that cannot influence
which items are proposed, and the added sequence length dilutes the SID
prediction. Suffix~Head and Suffix~MultiTower are within $0.01$--$0.05$ of each
other on every metric, so relaxing the single-label assumption into independent
per-behavior towers yields no measurable gain; the shared decoder hidden state
already carries what a behavior classifier can extract.

Reranking sharpens the top of the list but not the list itself. For
Suffix~Head, HR@1 rises from $4.11$ (beam) to $4.88$ (+rerank) and Recall@1
from $2.50$ to $2.99$, while every Top-$512$ column is identical before and
after. The same holds for Suffix~MultiTower ($4.44\!\to\!4.87$ HR@1). This is
the defining limit of output-side behavior modeling: the candidate set is
generated before the behavior signal is consulted, so the label can only reorder
a fixed beam and cannot pull behavior-relevant items into it. Placing the
behavior instruction on the input side, as in our main model, is what lets the
target behavior shape the candidate set during generation rather than after it.

\section{Post-training Algorithm}
\label{app:algorithm}

\begin{figure}
    \includegraphics[width=1.0\linewidth]{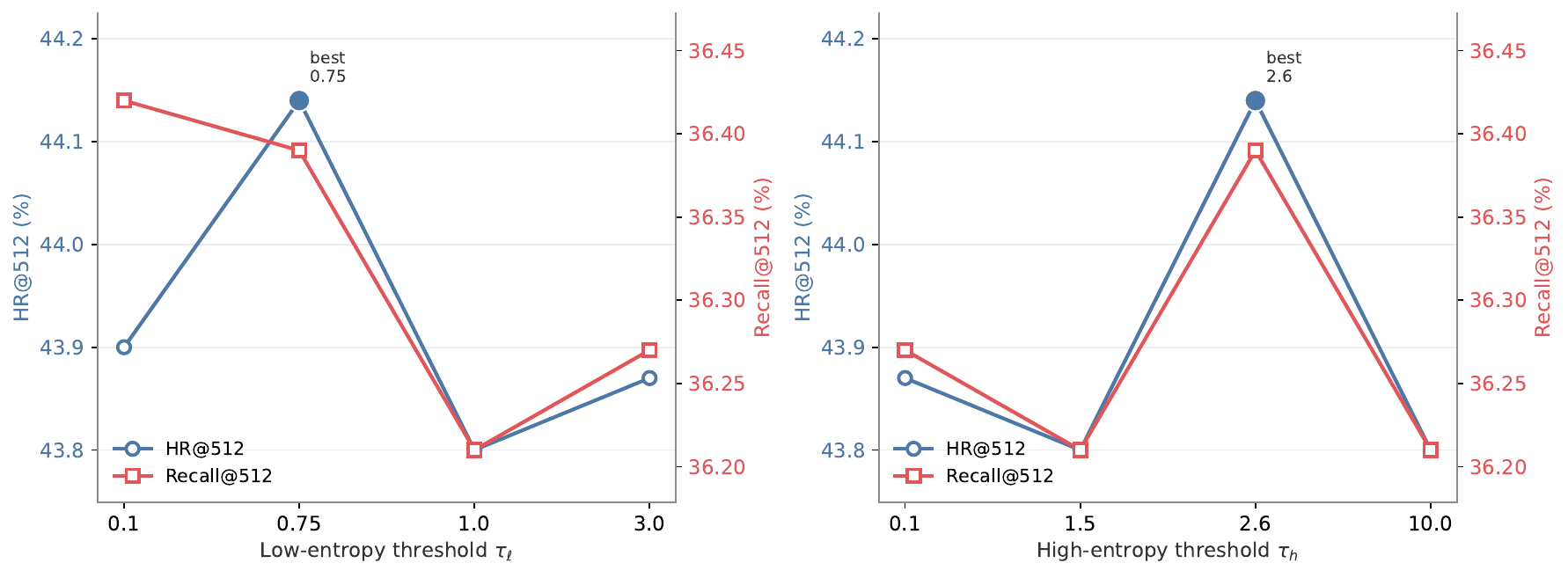}
    \caption{Sensitivity of post-training to the entropy gates
    $(\tau_l,\tau_h)$ of the privileged distillation,
    reported as HR@512 (blue, left axis) and Recall@512 (red, right axis).
    (a) Fix the best high-entropy threshold and sweep the low-entropy
    threshold $\tau_l\!\in\!\{0.1, 0.75, 1.0, 3.0\}$; the peak sits at
    $\tau_l\!=\!0.75$. (b) Fix the best low-entropy threshold and sweep
    the high-entropy threshold $\tau_h\!\in\!\{0.1, 1.5, 2.6, 10.0\}$; the
    peak sits at $\tau_h\!=\!2.6$. Both axes share the same scale across
    panels for direct comparison.}
    \label{fig:entropy-thresholds}
\end{figure}

The verifiable reward is anchored on the real behavior target but is sparse: most
rolled-out items do not exactly reproduce the gold SID prefix. Algorithm~\ref{alg:post} summarizes one post-training step. Because
verifiable hits are sparse across the $G$ rolled-out trajectories, we
concentrate the reinforcement signal on the highest-scoring segment
$k=\arg\max_i R_i$,
avoiding the noisy gradients that arise when low-reward trajectories are
mixed in. Along that segment, the teacher entropy $H^{T}_{t}$ routes each
token to one of two branches: low-entropy positions receive advantage-
weighted self-distillation $\mathcal{L}_{\mathrm{SD}}$, where a clear
teacher preference exists, while high-entropy positions receive forward
KL $\mathcal{L}_{\mathrm{FKL}}$, which transfers the teacher's uncertainty
structure and prevents premature mode collapse. A behavior-weighted SFT
anchor $\mathcal{L}_{\mathrm{SFT}}$ runs in parallel to keep the policy
close to the pre-training distribution. In this paper, we set $\lambda=0.1$, $\beta=0.01$, $\zeta=0.01$, $\tau_l=0.75$, and $\tau_h=2.6$. Figure~\ref{fig:entropy-thresholds} shows the performance of different entropy thresholds.

\begin{algorithm*}[t]
\caption{One post-training step of EA-TOSD: verifiable-reward best-trajectory optimization, entropy-aware privileged distillation, and a supervised anchor.}
\label{alg:post}
\textbf{Require:} minibatch $\mathcal{B}\subset\mathcal{D}$ of contexts $x$ with listwise gold $Y^{\star}=(y^{\star}_{1},\dots,y^{\star}_{L})$; backbone $\pi_\theta$; rollout group size $G$; geometric decay $\gamma$; teacher-entropy gates $\tau_l\!\le\!\tau_h$; objective weights $\lambda,\beta,\zeta,\lambda_{\mathrm{sft}}$; learning rate $\eta$; constant $\epsilon$.
\begin{algorithmic}[1]
\STATE Let $\pi_S(\cdot\mid X)$ and $\pi_T(\cdot\mid X,F)$ be the same model $\pi_\theta$ under student and teacher conditioning, where $F$ is the future target privileged prefix.
\FOR{all $x\in\mathcal{B}$}
  \STATE Sample $G$ on-policy trajectories $\{\hat{Y}_i\}_{i=1}^{G}\sim\pi_S(\cdot\mid X)$ 
  \STATE Compute the geometrically weighted per-trajectory verifiable reward and select the best-scoring segment:
  \begin{equation*}
   \quad\quad\quad\quad\quad\quad \omega_t=\frac{\gamma^{\,t-1}}{\sum_{j=1}^{L}\gamma^{\,j-1}},\quad
    R_i=\sum_{t=1}^{L}\omega_t\,\mathbb{I}\!\left[\hat{y}_{i,t}=y^{\star}_{t}\right],\quad
    k=\arg\max_{i\in\{1,\dots,G\}} R_i
  \end{equation*}
  \STATE Form the verifiable-reward objective on the selected trajectory:
  \begin{equation*}
    \mathcal{L}_{\mathrm{VR}}(\theta)=-R_k\sum_{t=1}^{L}\log \pi_S\!\left(\hat{y}_{k,t}\mid \hat{y}_{k,<t},X\right)
  \end{equation*}
  \STATE Compute the teacher entropy along segment $k$ and the normalized low/high-entropy gates:
  \begin{equation*}
   \quad\quad\quad\quad H^{T}_{t}=-\!\sum_{v\in\mathcal{V}}\pi_T(v\mid \hat{y}_{k,<t},X,F)\log \pi_T(v\mid \hat{y}_{k,<t},X,F),\quad
    \tilde{g}^{l}_{t}=\frac{\mathbb{I}(H^{T}_{t}<\tau_l)}{\sum_{j=1}^{L}\mathbb{I}(H^{T}_{j}<\tau_l)+\epsilon},\quad
    \tilde{g}^{h}_{t}=\frac{\mathbb{I}(H^{T}_{t}>\tau_h)}{\sum_{j=1}^{L}\mathbb{I}(H^{T}_{j}>\tau_h)+\epsilon}
  \end{equation*}
  \STATE Entropy-aware privileged distillation with privilege advantage $\mathcal{A}_t=\log \pi_T(\hat{y}_{k,t}\mid \hat{y}_{k,<t},X,F)-\log \pi_S(\hat{y}_{k,t}\mid \hat{y}_{k,<t},X)$:
  \begin{equation*}
  \begin{aligned}
    \quad\quad \mathcal{L}_{\mathrm{SD}}(\theta) &= -\sum_{t=1}^{L}\tilde{g}^{l}_{t}\,\operatorname{sg}\!\left(\mathcal{A}_t\right)\,\log \pi_S\!\left(\hat{y}_{k,t}\mid \hat{y}_{k,<t},X\right),\\
    \quad\quad \mathcal{L}_{\mathrm{FKL}}(\theta) &= \sum_{t=1}^{L}\tilde{g}^{h}_{t}\,\mathrm{KL}\!\left(\pi_T(\cdot\mid \hat{y}_{k,<t},X,F)\;\|\;\pi_S(\cdot\mid \hat{y}_{k,<t},X)\right)
  \end{aligned}
  \end{equation*}
  \STATE Form the supervised anchor $\mathcal{L}_{\mathrm{SFT}}(\theta)$: behavior-weighted next-token cross-entropy on the gold targets $y^{\star}$
  \STATE $\mathcal{L}_{EA-TOSD}(\theta)\leftarrow\lambda\,\mathcal{L}_{\mathrm{VR}}(\theta)+\beta\,\mathcal{L}_{\mathrm{SD}}(\theta)+\zeta\,\mathcal{L}_{\mathrm{FKL}}(\theta)+\lambda_{\mathrm{sft}}\,\mathcal{L}_{\mathrm{SFT}}(\theta)$
\ENDFOR
\STATE Compute $\mathcal{L}(\theta)\leftarrow\dfrac{1}{|\mathcal{B}|}\sum_{x\in\mathcal{B}}\mathcal{L}(\theta)$ and update $\theta\leftarrow\theta-\eta\,\nabla_\theta\mathcal{L}(\theta)$
\end{algorithmic}
\end{algorithm*}

\section{Privileged Teacher}
\label{app:teacher}

This section details the privileged teacher of Section~\ref{sec:method}: how the
future-target prefix $F$ is built, the teacher variants set by the behavior
threshold, and a formal statement of the sparse-reward gradient problem that the
distillation is designed to fill.

\subsection{Future-Target Prefix Construction}
\label{app:teacher_prefix}

The teacher shares the student backbone and differs only by an extra decoder
prefix $F$ that encodes the SIDs of the user's near-future targets. Let
the current sample be taken at timestamp $t_0$, and let the logged targets carry
timestamps $\{t_j\}$ and behaviors $\{b_j\}$. We order behaviors by the same
priority used throughout the pipeline,
\begin{equation}
  \rho(\text{any})\!=\!0 \prec \rho(\text{click})\!=\!1 \prec
  \rho(\text{cart})\!=\!2 \prec \rho(\text{order})\!=\!3,
\end{equation}
and fix a threshold $\rho_{\min}$ for the teacher variant. Scanning the targets in
time order, we keep a target $j$ as a future target when it lies strictly in the
future and clears the threshold,
\begin{equation}
  t_j > t_0 \quad\text{and}\quad \rho(b_j) \ge \rho_{\min}.
\end{equation}
Each kept target is mapped to its $c$ SID tokens ($c\!=\!3$, the codebook
depth), targets that contain a padding code are skipped, and we collect at most
$m$ future targets (default $m\!=\!2$). The result is truncated or padded to the
fixed length $m\cdot c$, giving the token block
\begin{equation}
  F = \big[\,\mathrm{sid}(j_1),\,\mathrm{sid}(j_2),\,\dots\,\big]\in\mathbb{Z}^{m\cdot c}.
\end{equation}
The block is embedded through the \emph{same} shared decoder embedding table used
for the behavior instruction and the SID tokens, and concatenated after the
student prefix, so the teacher decodes from $[\textsc{bos}, I_s, I_r, I_b, F]$ while
the student decodes from $[\textsc{bos}, I_s, I_r, I_b]$. The teacher prefix is thus
$m\cdot c$ embeddings longer than the student prefix; the loss routines align the
two by this fixed offset. Because the teacher is the policy itself with a short
look-ahead, the student and teacher distributions stay close and the distillation
is on-policy, and no separate or larger teacher model is required.

\subsection{Teacher Variants}
\label{app:teacher_variant}

The threshold $\rho_{\min}$ on future targets controls how dense and how clean the
teacher's target pool is, and defines the teacher variants ablated in the
experiments (Table~\ref{tab:teacher_app}). A permissive threshold admits exposed
items and yields a dense but noisy pool; a strict threshold keeps only orders and
yields a sparse, high-precision one. The ordering matches the behavior-value scale
that sets the instruction priority and the pre-training loss weights, so a single
behavior hierarchy runs through the instruction, the loss, and the teacher.

\begin{table}[h]
\centering
\footnotesize
\setlength{\tabcolsep}{5pt}
\caption{Teacher variants set by the behavior threshold $\rho_{\min}$ on future
targets. A lower threshold admits weaker behaviors and produces a denser, noisier
look-ahead pool.}
\begin{tabular}{@{}llcc@{}}

\toprule
Teacher variant & Behaviors kept & $\rho_{\min}$ & Pool \\
\midrule
Add exposure (default) & expo, clk, cart, ord & $0$ & dense / noisy \\
Add click              & clk, cart, ord       & $1$ & medium \\
Order only             & ord                  & $3$ & sparse / precise \\
\bottomrule
\end{tabular}

\label{tab:teacher_app}
\end{table}

\section{Hyperparameter Sensitivity Analysis}
\label{app:hyper}

We analyze how sensitive pre-training is to the strength of the behavior-value
weighting introduced in the behavior-aware training scheme. The default weights
are $w_b=\{1.2,1.5,2.0\}$ for click, cart, and order. To probe the effect of the
\emph{contrast} between behaviors rather than the absolute scale, we introduce a
gap-scaling factor $k$ that scales each behavior's offset from the click weight:
order and cart are set to $1.2+k\,(w_b-1.2)$ while click is fixed at $1.2$. Thus
$k{=}1$ recovers the default weights, $k{<}1$ flattens the differences toward
uniform weighting, and $k{>}1$ sharpens the contrast between high-value and
low-value behaviors. All variants share the behavior instruction $I_b$ and
exposure filtering, and differ only in $k$; every model is trained to the same
$\approx\!295$k steps under identical data and optimization.

\begin{figure}[t]
\centering
\includegraphics[width=0.72\columnwidth]{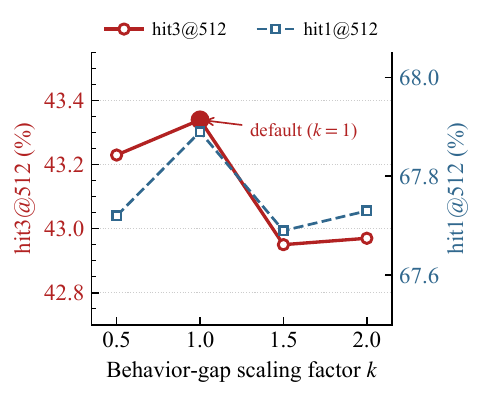}
\caption{Sensitivity of pre-training to the behavior-gap scaling factor $k$
around the default behavior weights $\{1.2,1.5,2.0\}$ for click/cart/order.
Layer-1 and layer-3 hit rate at Top-512 both peak at the default $k{=}1$;
flattening ($k{<}1$) or sharpening ($k{>}1$) the behavior contrast degrades
retrieval. hit$k$@512 denotes the layer-$k$ hit rate at Top-512.}
\label{fig:wbk}
\end{figure}

Figure~\ref{fig:wbk} reports the layer-1 and layer-3 hit rate at Top-512 as a
function of $k$. Both metrics peak at the default $k{=}1$: flattening the
behavior gap ($k{=}0.5$) and sharpening it ($k{=}1.5, 2.0$) each degrade the hit
rate, with the larger contrasts hurting the most (hit3@512 drops from $43.34$ at
$k{=}1$ to $42.95$ at $k{=}1.5$). This indicates that the default behavior
weights already lie in a favorable region: the model benefits from
distinguishing high-value orders from clicks, but pushing the contrast beyond the
default over-emphasizes rare high-value behaviors at the cost of overall
retrieval quality. We therefore keep $k{=}1$ throughout.

\section{Additional Experimental Setup}
\label{app:setup}

\subsection{Dataset and Item Tokenizer}
\label{app:setup_data}

We use daily interaction logs from a large industrial e-commerce platform in
Parquet format. Training spans one month (2026-03-25 to 2026-04-25) and
evaluation uses the next day (2026-04-26). Each example is a user context---short
and lifelong behavior sequences together with profile and contextual
features---paired with a listwise group of engaged items, each labeled by one of
four behavior types (exposure, click, add-to-cart, order). Items are tokenized
with a three-level residual-quantization SID
tokenizer~\cite{tiger,lee2022rqvae}: every item maps to an ordered triple
$(s^1,s^2,s^3)$ over three codebooks of size $8192$ each, so decoding one item
takes three steps. Table~\ref{tab:data} lists the key data settings.

\begin{table}[h]
\centering
\small
\caption{Dataset and item-tokenizer configuration.}
\label{tab:data}
\begin{tabular}{@{}ll@{}}
\toprule
Field & Value \\
\midrule
Domain            & Industrial e-commerce (proprietary) \\
Format            & Parquet, random seed $1$ \\
Train span        & 2026-03-25 to 2026-04-25 \\
Eval day          & 2026-04-26 \\
Behavior types    & exposure, click, add-to-cart, order \\
Short seq.\ length    & 20 \\
Lifelong seq.\ length & 256 \\
Listwise group size $G$ & $16$--$20$ \\
SID codebooks     & $3 \times 8192$ (residual quant.) \\
\bottomrule
\end{tabular}
\end{table}

\subsection{Model and Optimization}
\label{app:setup_model}

The backbone is an encoder--decoder transformer with an MoE decoder. Table~\ref{tab:model} summarizes
the architecture and the optimization schedule shared by all pre-training runs.
The offline studies use a 0.7B-parameter configuration; the online deployment
(Section~\ref{sec:online-ab}) scales the same architecture to a 3B-parameter,
1B-activated (3B-A1B) MoE. Post-training reuses this backbone and decoding format
and adds no inference-time module.

\begin{table}[h]
\centering
\small
\caption{Backbone architecture and optimization settings (shared across runs).}
\label{tab:model}
\begin{tabular}{@{}ll@{}}
\toprule
Field & Value \\
\midrule
$d_{\text{model}}$ / $d_{\text{ff}}$ & $2048$ / $2048$ \\
Encoder / decoder layers & $4$ / $8$ \\
Attention heads / $d_{kv}$ & $8$ / $128$ \\
Decoder experts (MoE) & $8$ \\
Tied embeddings & yes \\
Vocabulary size & $25{,}000$ \\
Optimizer & AdamW, $\text{lr}=1\times10^{-4}$ \\
LR schedule & WarmupCosine, warmup $20$k, total $1$M \\
Global batch size & $1024 \times 16 = 16{,}384$ \\
ZeRO stage & $1$ (overlap comm., $50$MB bucket) \\
Precision & bf16 \\
Training steps & ${\approx}295$k (EOS-terminated) \\
Decoding & beam $512$, length $4$, greedy \\
Eval sample count & $102{,}400$ \\
\bottomrule
\end{tabular}
\end{table}

\begin{table*}[p]
\centering
\small
\setlength{\tabcolsep}{5pt}
\caption{The formal definition of notations used throughout the paper.}
\label{tab:notation}
\begin{tabular}{@{}ll@{}}
\toprule
Symbol & Meaning \\
\midrule
\multicolumn{2}{@{}l}{\emph{Inputs and instructions}} \\
$X$ & user interaction context (behavior sequences, profile, context) \\
$I_s, I_r$ & scenario instruction and contextual reasoning instruction \\
$I_b$ & behavior instruction, $I_b=\psi(\mathrm{Emb}(b))$ \\
$b$ & target behavior, one of click, cart, order (exposure at train time) \\
$w_b$ & behavior-value weight for behavior $b$ \\
$F$ & privileged future-target prefix (teacher only) \\
\midrule
\multicolumn{2}{@{}l}{\emph{Items, tokens, and sequences}} \\
$(s^1,s^2,s^3)$ & SID triple of one item, over three codebooks \\
$N$ & number of items in a listwise recommendation \\
$L$ & number of SID decode steps, $L=3N$ \\
$t$ & decode-step (SID-token) index, $t=1,\dots,L$ \\
$y_t$ & SID token generated at step $t$ \\
$Y$ & SID-token sequence $(y_1,\dots,y_L)$; target in the pre-training objective \\
$\hat{Y}$ & generated candidate sequence (scored by the external reward)\\
$\hat{y}$ & generated SID token of candidate sequence\\
$Y^\star$ & ground-truth SID-token sequence \\
$Y_i, y_{i,t}$ & $i$-th sampled sequence and its token at step $t$ \\
$Y_{<t}$ & prefix of $Y$ before step $t$ \\
$\mathcal{V}$ & SID vocabulary; $v\in\mathcal{V}$ a candidate token \\
$\mathcal{D}$ & training dataset \\
$\mathcal{T}$ & pre-training set of (SID token, behavior) pairs \\
\midrule
\multicolumn{2}{@{}l}{\emph{Policies and post-training}} \\
$\pi_S, \pi_T$ & student and privileged-teacher policies \\
$G$ & Number of sampled trajectories \\
$h_{i,t}$ & per-step hit indicator, $\mathbb{I}[y_{i,t}=y^\star_t]$ \\
$\omega_t, \gamma$ & geometric step weight and its decay \\
$R_i, R_k$ & verifiable score of sequence $i$; best in the group \\
$\mathcal{A}_t$ & privilege advantage at step $t$ \\
$H_t^T$ & teacher predictive entropy at step $t$ \\
$\tau_l, \tau_h$ & low- and high-entropy gate thresholds \\
$g_t^l, g_t^h$ & low- and high-entropy gates; $\widetilde g$ normalized \\
$\mathcal{L}_{\text{VR}},\mathcal{L}_{\text{SD}},\mathcal{L}_{\text{FKL}}$ & verifiable-reward, distillation, and forward-KL losses \\
$\lambda,\beta,\zeta$ & post-training weights on $\mathcal{L}_{\text{VR}},\mathcal{L}_{\text{SD}},\mathcal{L}_{\text{FKL}}$ \\
\bottomrule
\end{tabular}
\end{table*}

\section{Training Dynamics Visualization}
\label{app:dynamics}

We trace both training stages to confirm that the behavior signal helps
optimization without destabilizing it.

\begin{figure*}[t]
    \centering
    \includegraphics[width=0.8\textwidth]{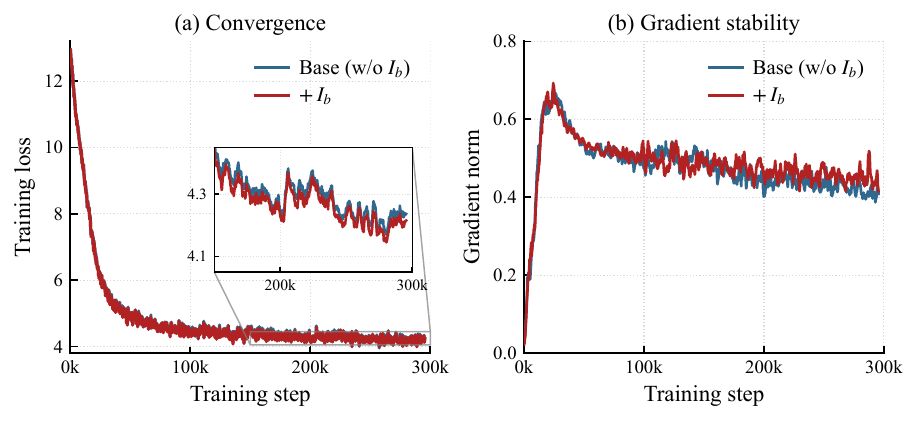}
    \caption{Pre-training dynamics over 295k steps. \textbf{(a)} Training loss for
the Base generator (w/o $I_b$) and $+\,I_b$; the inset over the converged tail
resolves the small, consistent margin the behavior instruction opens. \textbf{(b)}
Gradient norm for both runs, sharing the same warmup peak and settling into one
stable regime.}
    \Description{Two side-by-side line plots of pre-training. The left panel
    plots training loss against step for the Base generator and the +I_b run,
    both dropping from about 13 to just above 4, with a zoom inset over the last
    third of training showing the +I_b curve slightly below the Base curve. The
    right panel plots gradient norm against step, with both runs rising to a
    warmup peak near 0.65 and decaying to a stable band around 0.45.}
    \label{fig:pretrain-dynamics}
\end{figure*}

\paragraph{Pre-training.} Figure~\ref{fig:pretrain-dynamics} compares the two
configurations whose training logs we retain: the Base generator and the model
with the behavior instruction $I_b$. Both follow the same trajectory, dropping
from near $13$ to just above $4$ and flattening after roughly $30$k steps. The
instruction adds only a short decoder prefix, so it neither slows convergence nor
disturbs the gradient norm, which peaks together during warmup and decays into a
single stable band. The zoom inset shows the payoff: through the converged tail
$+\,I_b$ holds a small but consistent margin below Base, matching the retrieval
gain in Table~\ref{tab:preablation}. Conditioning on the target behavior therefore
lowers the objective it is trained on rather than merely reweighting outputs.

\begin{figure*}[t]
    \centering
    \includegraphics[width=0.86\textwidth]{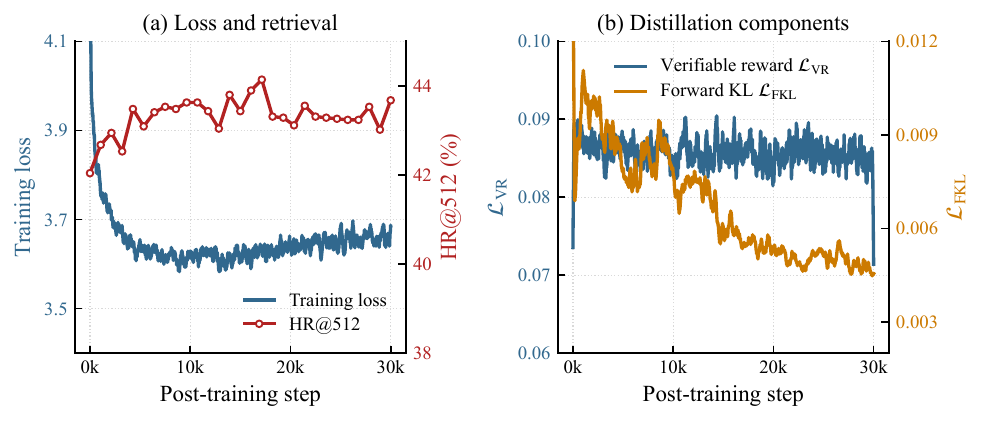}
    \caption{Post-training dynamics of EA-TOSD over 30k steps. \textbf{(a)} The
supervised training loss falls while HR@512 climbs to $44.1\%$. \textbf{(b)} The
two active distillation components, the verifiable-reward hit loss
$\mathcal{L}_{\mathrm{VR}}$ and the high-entropy forward-KL term
$\mathcal{L}_{\mathrm{FKL}}$, both settle into a stable regime.}
    \Description{Two side-by-side line plots of post-training. The left panel
    plots training loss and HR@512 on twin axes against step, with loss decreasing
    and HR@512 rising to about 44 percent. The right panel plots the
    verifiable-reward loss and the forward-KL loss on twin axes, both declining and
    stabilizing.}
    \label{fig:loss-hr512}
\end{figure*}

\paragraph{Post-training.} Figure~\ref{fig:loss-hr512} tracks the same backbone
under EA-TOSD. The supervised loss falls while HR@512 rises to $44.1\%$
(Figure~\ref{fig:loss-hr512}a), so the verifiable reward and privileged
distillation improve retrieval without the divergence that an unconstrained proxy
reward can cause. The two distillation terms converge in step
(Figure~\ref{fig:loss-hr512}b): the verifiable-reward hit loss and the
high-entropy forward-KL term both decline and flatten, and the supervised anchor
keeps the policy near the pre-training distribution, visible in the smooth loss
trace across the run.

\end{document}